# Client Monitoring Software: A Monitoring Tool for Greatleaf Land Inc.

Carlo H. Godoy Jr.[1], Jerico C. Torayno[1], Audrey Rose Abbey C. Magtarayo[1],
Mark Wilson J. Suarez[1], Armando Embile[1], Daven Christian O. Estopia[1]
[1]College of Industrial Technology,
[1]Technological University of the Philippines
Manila, Philippines
carlo.godoyjr@tup.edu.ph, jerico.torayno@tup.edu.ph, audreyroseabbey.magtarayo@tup.edu.ph
markwilson.suarez@tup.edu.ph, armando.embile@tup.edu.ph, davenchristian.estopia@tup.edu.ph

*Abstract*— **Monitoring typically supports greater analysis and allows for a lot deeper data collection on a Web browser level. Analysts may usually see the use of web-based monitoring software within an entire client context when it comes to client-side monitoring, on the other hand. In the case of Greatleaf Land Inc., their monitoring is somehow traditional. Traditional means that the paper method of monitoring is being utilized. When arranging piles of paper and afterwards forgetting about it, users lose track of where the information is situated. Another notable issue is that some information situating on a written on paper isn't easily can't be detected beyond skimming, which is a process normally being used to fast track records on a flipping pages, and sometimes there's also the fact that written records situating on papers can only situate in one location at a certain time unless the custodian has given up and made copies for everything. This study will check the feasibility of a web-based client monitoring software for Greatleaf Land Inc. By using ISO 25010, the study will determine if it is beneficial for Greatleaf Land Inc. to have an online monitoring software rather than settling to the traditional method that is currently being used by their company.**

I. INTRODUCTION AND BACKGROUND OF THE PROBLEM

In a Web browser level, Monitoring normally supports better analysis and enables a much richer data collection. On the other hand, client-side monitoring in a broad perspective, analysts normally can see the used of web-based monitoring software within an overall client context [1]. On the other hand if an organization is not using yet an online monitoring software, the method that is being used is searching manually through preserved records. Searching through preserved records normally demands for a large amount of time from man power labor as everybody knows that searching through a pile of records is not an easy task.

Another thing is that one problem of locating useful information in preserved records is known as document retrieval. There is a collection of documents on a variety of subjects, written by various writers at various times and with varying degrees of scope, detail, clarification, and precision, as well as a collection of individuals who, at various times and for various purposes, search for recorded information that may be found in any of the documents in this collection. Each when a person seeks information, the person will find certain documents in the set useful and others not so useful; the documents considered useful are important, while the others are not [2].

Aside from losing time in finding non-useful record, in the traditional method, users normally lose track of where information is located while arranging heaps of paper and then forgetting about it. Another notable issue is that some information situating on a written-on paper isn't easily can't be detected beyond skimming, which is a process normally being used to fast-track records on a flipping page, and sometimes there's also the fact that written records situating on papers can only situate in one location at a certain time unless the custodian has given up and produced several duplicates [3].

It takes a long time to file paper by hand. The person must not only organize and preserve the data, but the person must also locate the documents when it is being needed. Finding a file can take anything from minutes to hours, depending on how well organized the persons are. This can irritate both clients and employees. Workers are less productive since the workers have to squander too much time working with a paper file system [4]. This kind of event may eventually lead to workers having a burnout syndrome.

In line with this, everyone understands that if burnout is not prevented in advance, it might lead to burnout syndrome. Freudenberger coined the term "burnout" in the 1970s, describing it as a gradual loss of emotional energy and motivation. This was the very initial evidence of burnout syndrome which has a scientific basis. If in any case that evidence of burnout is found in diverse group of working people with a very distinct constellations of a very distinct working status, it can be considered that this kind of condition may be a distinction of burnout that normally occurs in occupations which demands connection that is social in nature. Burnout symptoms are thought to appear when a person's job expectations and available resources are out of sync [5]. All of these aforementioned issues normally result to pending account for commission release, hence causing a revenue lost for Greatleaf Land Inc.,

According to an interview with the company, pending accounts for commission release means that an agent's sale is on hold for a length of time owing to unmonitored missing paperwork, which could result in account cancellations and harm the agent's financial income. The researchers propose that

Greatleaf Land Inc. develop client monitoring software using a programmer tool called the Android Studio to serve like a software monitoring aid which will be a perceived remedy to the company's problem. The client monitoring software will benefit the organization because it will alleviate the load of the problems listed above. The software can be accessed by Servicing Agents, who can see, change, and delete the buyer's information, while the admin will only be responsible for the application's maintenance and system updates.

## II. OBJECTIVE, SCOPE AND LIMITATION OF THE STUDY

### A. Objective

The main goal of this developmental research is to come up with a very capable Client Monitoring Software for Great leaf Land Inc. using Android Studio as Software Monitoring Tool. To be detailed, the specific goal of the research are the following: (1) Development of mobile app thru Android Studio that will monitor the client's activities. The system has the following features: (a) View Buyer's Info; (b) Add Buyer's Info; (c) Edit Buyer's Info; and (d) Remove Buyer's Info; (2) Create an app which can help servicing agents in accessing information sheets anytime and anywhere; (3) Perform testing and recreate the app after comparing its performance with test cases based on the requirements; and (4) Validate the efficiency of the developed app using a Software Quality Matrix based on ISO 25010;

### B. Scope and Limitations

The study will focus in developing a client monitoring software that will help the Servicing Agents of Great leaf Land Inc. The application software will be used to monitor client activities such as account status and account details. The application can be accessed by the Servicing Agents where the buyer's info can be viewed, edited, and removed. The admin will handle the maintenance and system updates of the application. The application is only applicable on Android devices as of now but can be implemented on other devices if still plausible to do so. The research needs to be completed within the given deadline.

This application must be pilot tested by a Branch Manager and/or by a Servicing Agent to determine whether there is an issue or any additional functionality for the application that needs to be added.

## III. RELATED WORK

According to the Columbia University Press (2007), the forms that are normally used for client monitoring has been being referred to as the critical records, personal logs diaries and journals of the client [6]. Regardless of the term being used, it still gives accurate data that hard to get by normal way of capturing. The Journal of Health Administration applied the usage of client monitoring as a client monitoring for the treatment progress scale of every patient. According to the study of Schalock et al., (1993), some of the staff of the clinic normally gets their patient's TPS scales in a monthly basis in the form of a hardcopy format. On the other hand, by the time that the patients will be discharge from a hospital or a clinic, the graphed profile of a patient is being used to get a summarization of the TPS change scores. This is categorized per problem area and normally, it will stay in the hospital and will be retained as the permanent record of the client for the purpose of sake keeping and then will be retrieve when the time comes of their readmission. The digitalization of the TPS or in other words transforming it to an electronic copy as well as the unit cost system normally gives differentiation between the cost of the hospitalization and the change in the. These shows how this administration apply the client monitoring, the burden of paperwork's lessens [7].

According Fenstermacher, Client monitoring normally is performing to give information, in fact, gives data regarding the how's, when, why's and what artifacts were normally being utilized in the knowledge discovery. Thus, a user is able by extension, store, document, and analyze their procedure in the knowledge bank of the organization to make future groups from having to start from scratch when doing the same work. In this study the client monitoring greatly benefit them in storing and analyzing data [8]. Based on the client/server mechanism, a software monitoring system for distributed programs. It enables the detection of output bottlenecks caused by both hardware and software flaws. Unlike other methods, the proposed framework would not restrict the user to a single view of the program, but rather provides for real-time analysis of its actions. It will more clearly distinguish when a client/server program faces undue latency by revealing interactions with other systems, such as resource rivalry. It therefore fills the void left by the absence of automated management frameworks for client/server apps, and it represents a first move toward reducing customer dissatisfaction induced by excessively long response times.

The manager was asked to assess the Monitoring Software's usefulness as a project management mechanism by contrasting it with the methods and processes in operation on their project. Time Benefit, Cost Benefit, Quality Benefit, Information Benefit, Process Benefit as well as Overall Effectiveness were among some metrics used to evaluate effectiveness [9]. One of the most common tools today in project management is Customer Relationship Management (CRM) software. In this matter they test how effectively the monitoring system may benefit the company in over all aspects. In today's world, CRM is a rapidly growing business activity. It is used to supervise in a digital manner an organization's daily, weekly or monthly interactions and agreements with some of their potential or their loyal and existing customers. Normally, usage of CRM approach means to analyze information regarding the company's consumer background. It focuses on a method on which the organization will be able to retain their customers, which normally help in raising the sales of the organization. As a positive end point, the company's trading relationship with the one which whom their services are being offered will improve. The aim of this research is to find out how technology, organizational capacity, customer orientation, and customer knowledge management all affect CRM performance [10].

CRM is the strategic method of identifying and influencing the customers that a business will most profitably serve, as well as the interactions that a company has with these customers. The objective is to maximize the company's existing and potential consumer value [11]. CRM also refers to the use of sophisticated techniques and engineering to locate, acquire, and cultivate privileged customers in order to sustain long-term relationships. The method of creating CRM and the content of CRM are the two main streams of CRM literature [12]. It is also a technique in management which gives organizations to locate, acquire, and eventually decides to retain productive customers thru managing relationships. Customer Relationship Management became popular as markets transforming to be more competitive. The different positive change in the field of information technology (IT) and the increasing popularity in the process of customer familiarization serves as fundamental concept in the business field are CRM's major push in the market [13].

Seraj et al., [14] remarked that technology is rapidly evolving and spreading throughout the world, including to the most isolated locations. CRM also entails automating and improving customer-centric distribution, marketing, and service processes. It focuses not only on automating these procedures, but also on ensuring that front office systems boost customer retention, resulting in improved customer engagement, which has a direct impact on the company's bottom line [11].

IV. PROPOSED METHODOLOGY

The proponents' proposed system is an android application that will work as a client monitoring software. The suggested IDE for creating an android application is the Android Studio since it is free for anybody who wants to make a professional application and is based on JetBrains IntelliJ IDEA software. Comparing to Eclipse, its demo and beta version were better thus the reason why many Android developers are using it as their IDE from the start [15]. Different algorithms are also used to make the functionality suited to the needs of Great leaf land inc. The use of algorithms is helpful because it is increasingly useful in determining what knowledge is most valuable to us, which is a critical aspect of our involvement in public life. Search engines assist us in navigating large databases of information, as well as the entire internet. Recommendation algorithms compare our tastes to those of others, recommending new or overlooked cultural gems for us to discover. Algorithms are in control of the lives of many people [16].

In accordance with the following concepts discussed, theories that were scrutinize and all the results obtained from related literature, as well as studies from various sources, the conceptual model of the study needs to be developed, as illustrated in Figure 1.

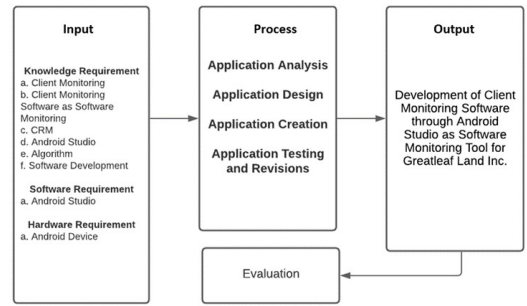

Figure 1: Conceptual Framework of the Client Monitoring Software for Greatleaf Land Inc.

A. Conceptual Framework

Figure 1 shows the conceptual framework Integrating the learnings from the different literature review as well as studying relevant systems, the researchers came up with a newly built system that will give Greatleaf Land Inc. a monitoring tool that is useful for their everyday needs.

B. Project Methodology Framework

This part contains the methods and procedures that will be performed in the gathering of data in the proposed framework. It presents the methods of research to be used, the instruments in the gathering of data, sampling technique in evaluating of system and software development model. The Client Monitoring Software (GLLI Agent's Client Monitoring) is a software monitoring tool used by the servicing agents of the Greatleaf Land Inc. It is an Android application that can help the servicing agents to monitor client activities such as account status and account details.

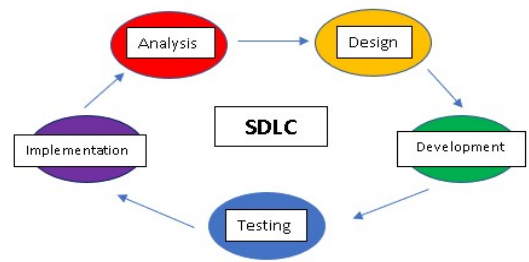

Figure 2. Prototyping Development Methodology

As shown in Figure 2, this study uses the Software Development Life Cycle (SDLC). The framework specified each task performed at each step of the software development process. It is made up of a comprehensive plan that explains how the system was created. The software development life cycle is made up of many well specified work phases namely Analysis, Design, Development, Testing, and Implementation.

C. Project Design

In developing the system, the software development model called the prototyping model will be applied as illustrated in Figure 2. The system will be divided into different modules. The different modules will be combined all together to form a complete system.

Once the system is done, it will be tested and evaluated. Different testing and evaluation procedures will be used. The following are the discussion of the procedures that will be done which includes the development, testing, and evaluation procedure.

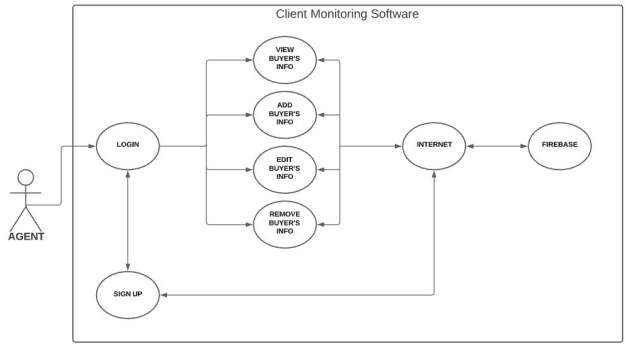

Figure 3. Use Case Model Diagram of the Client Monitoring Software for Greatleaf Land Inc.

Figure 3 shows the Use Case Model diagram of the system which illustrates how the user interacts with the system. The system has only one type of user which is the Agent. The Agent can login or sign up on the very first page on the application. Further in the application, once logged in they can view, add, edit, and remove the buyer's info sheet.
The application needs an internet connection to access the list of buyer's info sheet that are stored on a cloud-hosted database which is the Firebase Realtime Database. There are different types of databases out in the public just like the NoSQL cloud-based database which synchronizes data in real time among all clients and it has those offline features, a great example of this is Firebase Realtime Database which is being used on the the Client Monitoring Software for Greatleaf Land Inc. The data will be stored in JSON file in the Realtime database, which is all linked clients share a single instance, receiving automatic updates of the most recent data [17].

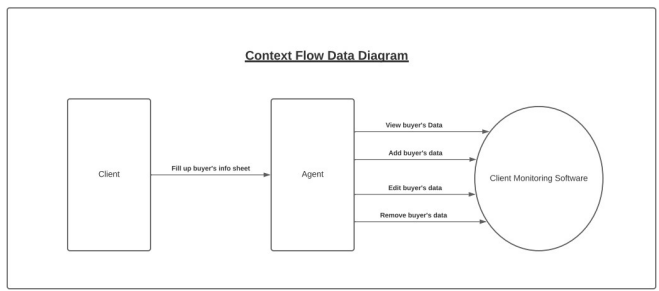

Figure 4. Context Flow Data Diagram of the Client Monitoring Software for Greatleaf Land Inc.

The Context Flow Data Diagram shown on Figure 3 shows the process of where the information came from and the one who receives the information and what can be done to the information gathered. The figure above has a very simple flow of data, starting from the client, the agent will ask for the buyer's info sheet from the client and if the client agrees the agent will service the filled-up buyer's info sheet through any type of internet messaging platform. The agent can add the information to the Client Monitoring Software in which can be viewed, edited, and removed.

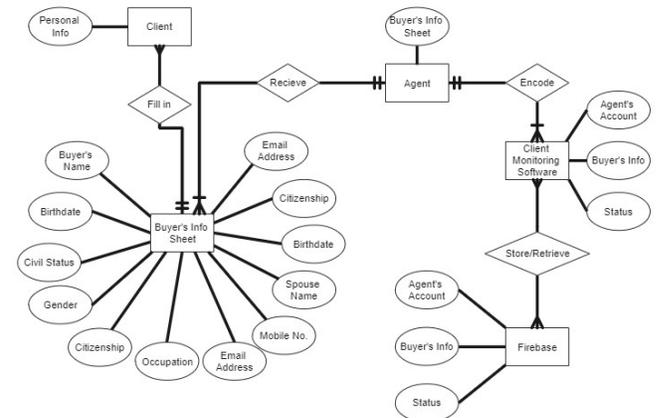

Figure 5. Entity-relationship diagram of the Client Monitoring Software for Greatleaf Land Inc.

In Figure 5, the Entity-relationship diagram describes the relationship of these entities, elements, and attributes. The Client can only have one and only one Buyer's Info Sheet to fill in while the Buyer's Info Sheet can be distributed to many Clients. The Agent can have one or many Buyers' Info Sheet while the Buyer's Info Sheet can only be given to one and only one Agent. On the Client Monitoring Software, the Agent can input one or many Buyer's Info Sheet, while there is one and only one Agent using their account to access Client Monitoring Software. The Client Monitoring Software can store the data on the Firebase by many and the Firebase can send the data to the Client Monitoring Software by many.

## V. OPERATION AND TESTING PROCEDURE

To ensure the system's quality, a series of tests were performed on each module in coordination with ISO25010, and the system was put through its paces by installing it in a live environment. The study used the client and researchers to run the app in their cellphone using Android Studio to briefly check if the system is working properly. The test plan was implemented after the development phase. Also, we run the app using Android Studio in every android OS version to see if it is working in majority of user's phone. Android Studio already has a feature that will give you an option to what version of android your developing application can run but we tested it to double check.

**Functionality Test.** The aim of a functionality test is to ensure that the application meets all the specifications and implements all of the functionalities defined in the functional requirements. Following the development process, the researcher took the following steps: (1) Prepared functionality test cases for each module; (2) Executed the test cases; (3) Recorded the test outputs; (4) Fixed all failed test cases; and (5) Re-executed failed test cases to verify that the test cases were resolved.

**Reliability Test.** This test was carried out to ensure that the device would correctly perform the stated functions under specific conditions for a set period of time. Live testing is used to ensure that the device is reliable. The device is put into action in a real-world setting and put through its paces. The following steps were performed: (1) Test the system in live environment; (2) Requested users composed of the client and researchers to install, use, and to test the system for a week; (3) Requested clients to forward the issues encountered back to the researcher; and (4) Gathered and tabulated the results.

**Compatibility Test.** The system's compatibility with various devices with different requirements, such as Android OS versions, hardware configurations, screen sizes, and network types or speed, was investigated using a compatibility test.

**Evaluation Procedure.** The following activities were conducted to evaluate the performance of the system and the outcome of the study: (1) Invited the client and researchers to be evaluators; (2) Downloaded and installed the system into their devices or run the app using android studio to discuss the features; (3) Requested the evaluator to rate the system using the ISO 25010 Evaluation Criteria for GLLI Monitoring System, and the rating scale of 1-5 where 5 is the highest and 1 is the lowest, as shown in Table 1.; (4) Collected and tabulated the data to get the overall mean, and the mean for each criterion; and (5) Interpreted the numerical value to get the equivalent descriptive rating using the interpretation shown in Table 2.

**Table 1.** *Rating Scale Used in the Evaluation*

| Rating | Interpretation |
|---|---|
| 5 | Excellent |
| 4 | Very Good |
| 3 | Good |
| 2 | Fair |

**Table 2.** *Scale Range and its Qualitative Interpretation*

| Range | Interpretation |
|---|---|
| 4.51 – 5.00 | Excellent |
| 3.51 – 4.50 | Very Good |
| 2.51 – 3.50 | Good |
| 1.51 – 2.50 | Fair |

VI. RESULTS AND DISCUSSION

**Project Description.** The GLLI Monitoring System is an application specifically made for Android Smartphone devices. The application's function is very straight forward since it was only meant to encode the Buyer's Info Sheet. The application uses Firebase Realtime Database to store the accounts and credentials, the first page of the application has a login and register function for a new Servicing Agent who intends to use the application. Once logged in, showing the second page of the application which consists of refresh list, add an entry, search bar, and a logout function. Tapping on the client's Buyer's Info Sheet brings us to the third page of the application where we have can view, edit details, confirm the update, and/or remove an entry.

**Project Structure.** The application is divided into four pages: (1) login and register; (2) the data list; (3) add an entry; and (4) the Buyer's Info Sheet page. First the login and register page for user authentication and registration. Second is the data list, mainly showing the list of names of the client of the Servicing Agents with additional functions which are refresh list, add an entry, search bar, and logout function. Third, on the add on entry it will ask for the credentials needed to be encoded on the application. Lastly, when tapped on the names on the Data List we enter the third page which is the Buyer's Info Sheet, this is where we view the details of the client, additionally we can edit and confirm the edit on the Buyer's Info Sheet or remove it.

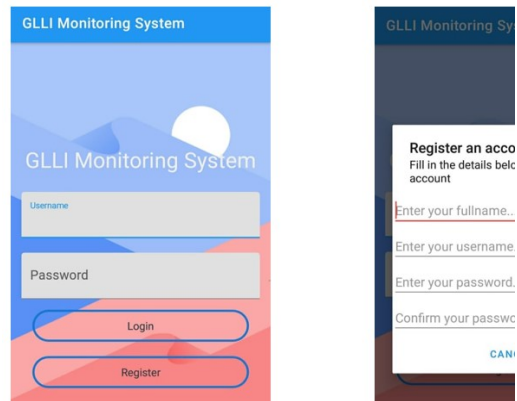
Figure 6.  Login and Register Page

**Login and Register Page.** The first page of the application serves as the user authentication and registration process of the application. The login only asks for the user's username and password. On the register, you will have to fill up the information need in order to create an account on the application.

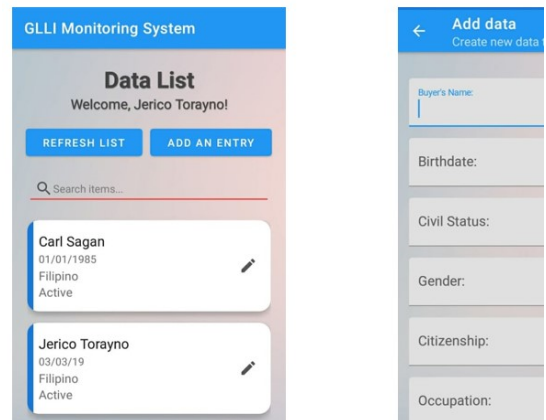
Figure 7.  Data List and Add Data Page

**Data List Page.** The second page of the application shows the name of the clients encoded by the Servicing Agents. The other functions on this page are the Refresh List where we refreshen the list by retrieving the information from the database, Add an Entry, a search bar to quickly search for a specific client, and the logout button to return to login and register page.

**Add an Entry Page.** For the third page of the application, the Add an Entry function allows the Servicing Agent to add more clients by encoding the credentials of their client. Filling up the credentials asked on this page isn't mandatory since it can be edited later on. Once the credentials of their client have been encoded, tapping on the confirm button will start to store the information to the database and base to the application.

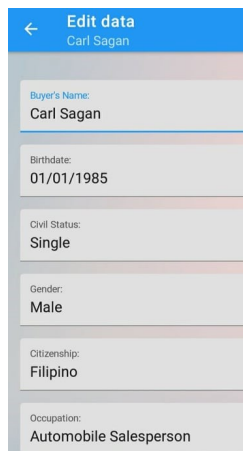

Figure 7. Edit Data Page

**Buyer's Info Sheet Page.** The very last page where everyone will get to see and edit the credentials of the clients, by clicking the confirm button we can rewrite the information that was saved on the database. The Remove this Entry function/button can delete the Buyer's Info sheet we are in, pressing the button will show the Delete Item warning message to avoid removing the Buyer's Info Sheet accidentally.

The GLLI Monitoring System went through a series of testing trying to cover every possible way to see the problems and improvements to be made to assess its functionality.

**Android Versions.** The application was tested on different Android devices with different Android versions. The Table 3 shows the result of the compatibility test that was conducted. The application was compatible to Android Smartphones with Android 8.0, 8.1, 9.0, 10.0, and 11.0 version. The application is also compatible within 5.2 to 6.3 inches of screen sizes.

**Table 3.** *GLLI Monitoring System Compatibility Test Result*

| Environment | Results |
|---|---|
| **Android OS Versions** | |
| Android 8.0 (Oreo) | User can access all pages with |
| Android 8.1 (Oreo) | User can access all pages with |
| Android 9.0 (Pie) | User can access all pages with |
| Android 10.0 (Q) | User can access all pages with |
| Android 11.0 (R) | User can access all pages with |
| **Screen Sizes** | |
| 5.2, 1080 x 1920, 420dpi | User can access all pages with |
| 5.5, 1080 x 2160, 440dpi | User can access all pages with |
| 5.5, 1440 x 2560, 560dpi | User can access all pages with |
| 5.6, 1080 x 2220, 440dpi | User can access all pages with |
| 5.7, 1440 x 2560, 560dpi | User can access all pages with |
| 5.7, 1080 x 2280, 440dpi | User can access all pages with |
| 5.8, 1080 x 2340, 440dpi | User can access all pages with |
| 6.0, 1440 x 2560, 560dpi | User can access all pages with |

**Evaluation Results.** The developed project was evaluated using an instrument with seven criteria: (1) functionality; (2) reliability; (3) compatibility; (4) usability; (5) performance efficiency; (6) security; and (7) portability. The summary of the results is shown in Table 4.

**Table 4.** *GLLI Monitoring System Evaluation Results using ISO25010*

| Criteria | Mean (X) | Descrip |
|---|---|---|
| Functionality | 4.63 | Ex |
| Reliability | 3 | G |
| Compatibility | 5 | Ex |
| Usability | 4.33 | Ver |
| Performance Efficiency | 3.5 | G |
| Security | 3.67 | Ver |
| Portability | 4.75 | Ex |

Based on the result of the evaluation, criterion that got the highest mean score is the compatibility. The compatibility criteria attained a mean of 5 with a descriptive rating of "Excellent". This means that the project was highly successful in terms of compatibility on Android Smartphone devices. Due to the application made specifically for Android Smartphone devices starting from Android 8.0 (Oreo) version and onwards made it compatible for these devices. The Servicing Agent can still see the encoded information even if they use a different Android Smartphone since the application uses Firebase Realtime Database in which they can easily retrieve the information.

Meanwhile, the reliability criteria got the lowest mean. It got a score of 3 with a descriptive rating of "Good". Despite it not having a problem with using different functions under Wi-Fi or mobile data, it does not work offline and if there is a software fault most likely it would not function completely as intended. The application has met the desired functionality requested by our client. It could encode the Buyer's Info Sheet and all its functions which are the login, register, logout, refresh list, viewing, add an entry, remove this entry, and edit data are working as intended. Thus, it had a score of 4.63 in functionality with a descriptive rating of "Excellent".

The application can perform the tasks it was intended for, and that the application was user friendly. The UI of the application was pleasant for the eyes so it would not be irritating to the eyes of the user through prolong use. The application was only given and installed to Servicing Agents of the GLLI. Thus, it had a score of 4.33 in usability with a descriptive rating of "Very Good". In terms of performance efficiency, the application loads fast when opened and operated. This will benefit the Servicing Agents in terms of efficiency since it shortens the time needed to wait for the application to load. Such other factors though like the slowing down of the application due to the numbers of Buyer's Info Sheet and the maximum limit of what can be encoded are undecided. Thus, it had a score of 3.5 with a descriptive rating of "Good".

The application is only given to Servicing Agents and information inside the application can be accessed only by the admin and Servicing Agents. Using Firebase as its database makes it secure through authentication of users and writing security rules. The downside though is that the information encoded on the application can be seen by other Servicing Agents. Thus, it had a score of 3.67 in security with a descriptive rating of "Very Good". The application would still be functional even with the evolving of hardware and newer software updates on Android Smartphone devices. The application has a low memory requirement that of 5.2 MB to install and it had replaced the use of traditional paper filings and Microsoft Excel for the Servicing Agent of GLLI. Thus, it had a score of 4.75 in portability with a descriptive rating of "Excellent".

In terms of the overall result, the Servicing Agent who had tested the application is pleased with the outcome and the capabilities of GLLI Monitoring System and had a score of 4.08 with a descriptive rating of "Very Good".

## CONCLUSION AND FUTURE WORKS

The system application, GLLI Monitoring System was designed based on the functionality and capabilities requested by the client. The application has a login and registration page for security reason and registration. It can encode the Buyer's Info Sheet which can be viewed on the Data List and when the name of the client is tapped it further expands to show the other credentials of that client. To easily find the clients, there is a search bar on the Data List. Additionally, the application can remove and edit previous encoded Buyer's Info Sheet and a logout button. After the evaluation, the application had an overall mean score of 4.08 with a descriptive rating of "Very Good" from the client.

Through various of studies and tests, researchers are able to come up with the monitoring system that can lessen the manpower of the employees of the Real Estate, Greatleaf Land Inc. By the guidance of the objectives created at the first part of the study, the following were obtained; the software was able to view, add, edit and remove buyer's information. The aim for it to be in a mobile device, wherein the employee can bring it anywhere and anytime, were also achieved. Lastly, the software has been tested by both the researchers and the client and both were satisfied by the outcome of the software.

Given these points, the researchers conclude that the software created was able to ease manpower and assist the employees of Greatleaf Land Inc. The following recommendations were drawn based in the summary of findings, conclusions, and comments of the evaluators and panelists: (1) Add choices for Civil Status (e.g., single, married, widowed, etc.); (2) Add choices for Gender (e.g., Male or Female); and (3) The access of information should be per Servicing Agent.

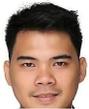 Mr. Carlo H. Godoy Jr.,NSE is a former Escalations Manager at Novartis Pharmaceutical and Support Analyst for SQL at Human Edge Software Philippines. Currently he is a Computer Programmer/Sail Plan Manager at the office N6 of the Philippine Navy. He is also a Research Scholar at the Technological University of the Philippines taking up Masters in Information Technology specializing in studies related to Educational Technology using Emerging Technologies like Augmented Reality.

Orchid ID: https://orcid.org/0000-0002-7701-8036

Web of Science ResearcherID: AAO-2785-2020